# RISC-V Toolchain and Agile Development-based Open-source Neuromorphic Processor

Jiulong Wang
Institute of Computing Technology, Chinese Academy of Sciences

Ruopu Wu
Beijing University of Posts and Telecommunications

Guokai Chen
Institute of Computing Technology, Chinese Academy of Sciences

Xuhao Chen
ShanghaiTech University

Jixiang Zong
University of Chinese Academy of Sciences

Boran Liu
Institute of Computing Technology, Chinese Academy of Sciences

Di Zhao*
Institute of Computing Technology, Chinese Academy of Sciences

## ABSTRACT

In recent decades, neuromorphic computing aiming to imitate brains' behaviors has been developed in various fields of computer science. The Artificial Neural Network (ANN) is an important concept in Artificial Intelligence (AI). It is utilized in recognition and classification. To explore a better way to simulate obtained brain behaviors, which is fast and energy-efficient, on hardware, researchers need an advanced method such as neuromorphic computing. In this case, Spiking Neural Network (SNN) becomes an optimal choice in hardware implementation. Recent works are focusing on accelerating SNN computing. However, most accelerator solutions are based on CPU-accelerator architecture which is energy-inefficient due to the complex control flows in this structure.

This paper proposes Wenquxing 22A, a low-power neuromorphic processor that combines general-purpose CPU functions and SNN to efficiently compute it with RISC-V SNN extension instructions. The main idea of Wenquxing 22A is to integrate the SNN calculation unit into the pipeline of a general-purpose CPU to achieve low-power computing with customized RISC-V SNN instructions 1.0 (RV-SNN 1.0), Streamlined Leaky Integrate-and-Fire (LIF) model, and the binary stochastic Spike-timing-dependent-plasticity (STDP). The source code of Wenquxing 22A is released online on Gitee[1] and GitHub[1]. We apply Wenquxing 22A to the recognition of the MNIST dataset to make a comparison with other SNN systems. Our experiment results show that Wenquxing 22A improves the energy expenses by 5.13 times over the accelerator solution, ODIN, with approximately classification accuracy, 85.00% for 3-bit ODIN online learning, and 91.91% for 1-bit Wenquxing 22A.

## 1 INTRODUCTION

With the development of bio-inspired computing, researchers are seeking a better way to imitate brain behaviors from neuroscience observation. The very first step is Artificial Neural Network (ANN), the important concept of Artificial Intelligence (AI) [22]. It ranges from Computer versions to sensory data processing. A large range of AI computing is realized by CPU or GPU. The hardware processes complex information which is difficult to process for human beings [20] in the Internet of Things (IoT) or Edging Computing. While researchers are pursuing higher accuracy of AI, the low-power performance of the neural network is also important. Thus, the Spiking Neural Network (SNN), classified as the third generation of neural network models, becomes an optimal choice. The SNN has more bio-inspired features to imitate the low-power computing of brains and it is computationally more powerful than other neural network models [1][2][23]. SNN uses the event-based model to imitate biological neurons with minimal energy. Recent works in SNN architecture design [6][7][8][9][10] show the potential of matching the accuracy of other non-spiking ANNs.

The traditional convolutional neurons require many operations, e.g., accumulating and multiplying the input

---

value with a weight. It is demonstrated that those arches have efficient efficient implementations on FPGAs through massive work [15]. Besides that, the event-based SNNs avoid lots of unnecessary calculations by only computing retrieved events. Outside sparse and dynamic events are assumed. This sparsity provides time- or order-based input data and excludes time-wasting memory access. Related works such as Intel Loihi [16], Tsinghua University Tianjic [17], and Frenkel's ODIN [14], have shown that processing event-based data by SNN can be efficient in both training and inference. However, these neuromorphic processors still have much power consumption. Thus, we propose an extremely low-power neuromorphic processor **Wenquxing 22A.**

In this work, we use a 9-stage inorder RISC-V processor, Nutshell designed by the University of Chinese Academy of Science OSCPU (Open-Source Chip Project by University) team [3], as our baseline and extend its execution unit with our self-designed RISC-V SNN extension instruction set. NutShell features RV64IMACSU instruction set and can run modern operating systems like Linux. This processor is also a base version of the XiangShan processor, a high-performance RISC-V open-source processor [21]. The first stable micro-architecture of XiangShan is called Yanqihu, which has been developed since June 2020. The current version of XiangShan, also known as Nanhu, is still under development.

Our customized RISC-V-based SNN instruction set has high computational granularity to prevent the pipeline from being stalled for a long time due to the execution of one instruction. We utilize the Leaky Integrate-and-Fire (LIF) model and the order-based Binary Stochastic STDP to compute event-based SNN. The neuron and synaptic models are both hardware-friendly and energy-efficient.

Here are our contributions to achieving low-power SNN computing:
- We design the customized SNN extension instruction set, RV-SNN 1.0, in high computational granularity based on RISC-V ISA;
- We streamline the standard LIF model to reduce the difficulty of computing and integrating the neuron model in our processor implementation.
- We modify the Binary Stochastic STDP to gear to the single cycle updating of synaptic weights.
- The source code of Wenquxing 22A will be released for further improvement and future works like taping out the neuromorphic chip.

In addition, to evaluate Wenquxing 22A, we implement our design to the Alveo U250 FPGA platform. Recognize the MNIST dataset using RV-SNN instructions to evaluate the power consumption and recognition accuracy of the SNN on Wenquxing 22A. We also apply another spiking accelerator solution, ODIN [14], on the same FPGA platform as the baseline of energy and accuracy comparison. The result turns out that Wenquxing 22A has a maximum recognition accuracy of 91.91% and improves the power consumption by 5.13 times over the accelerator solution of ODIN.

## 2 DESCRIPTION OF SOURCE CODE

This section introduces the released source code of Wenquxing 22A.

### 2.1 The Files of Wenquxing 22A

For a clear view of the overall directories of Wenquxing 22A, we list the related files as shown in **Figure 1**.

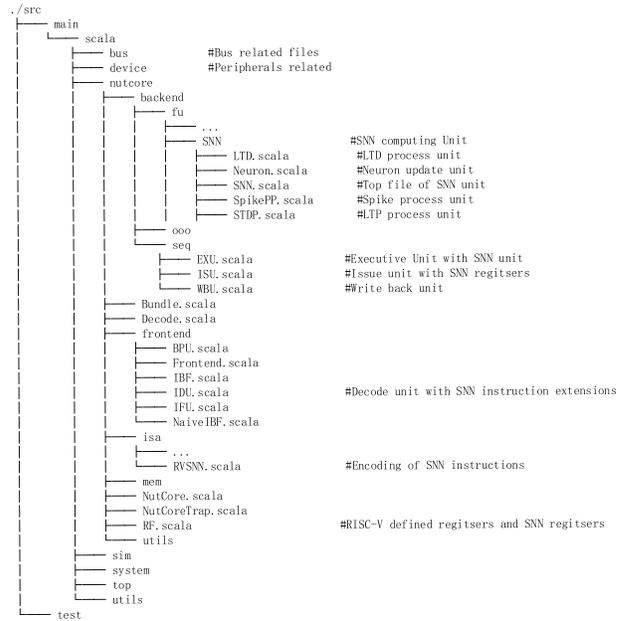

**Figure 1. Files of Wenquxing 22A**

### 2.2 The SNN Unit

Wenquxing 22A is adapted from Nutshell. The only difference is that Wenquxing 22A has an SNN Unit for spike process, neural update, and synaptic computing. This part is the core code of Wenquxing 22A, which is in the function unit (fu) directory.

The SNN Unit contains 4 more components which are the spike process unit, the Long-Term Depression Unit (LTD Unit), the Neuron Unit, and the Long-Term Potential (STDP Unit):

*Spike Process Unit-* This unit controls executing the AND operation of input spikes and synapses. Then the number of valid spikes will be counted and submitted to the next component.

*Neuron Unit-* The neuron Unit updates the current state of neurons depending on the number of valid spikes from the Spike Process Unit, the previous state, and the leakage voltage of the neuron.

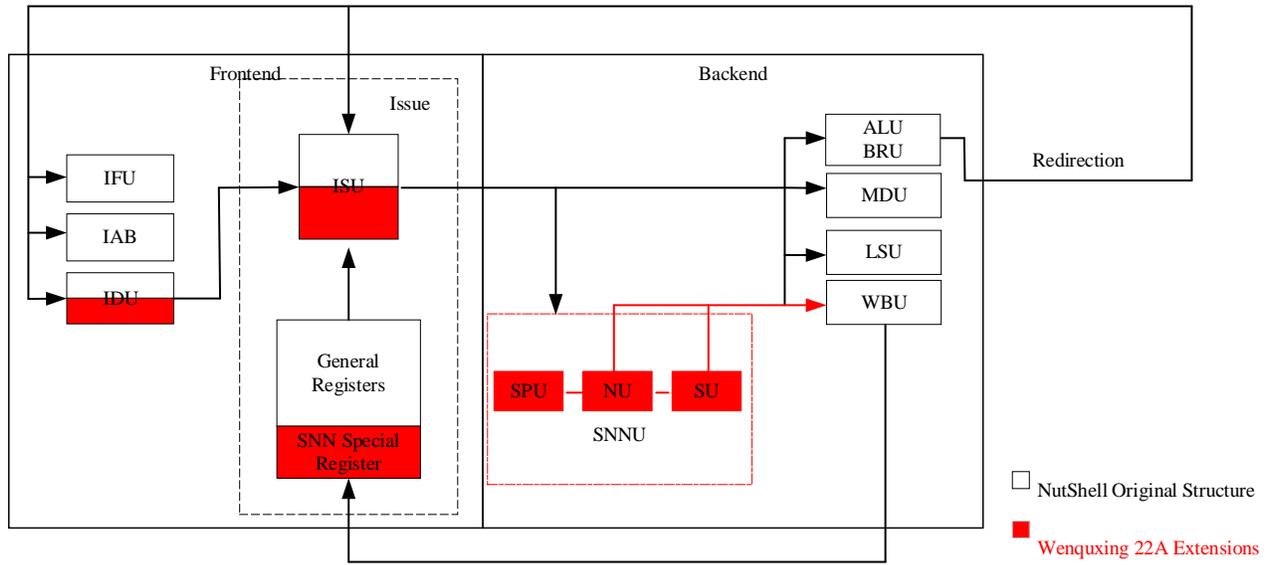

**Figure 2: Overall Architecture of Wenquxing 22A (Baseline: NutShell).**

*LTP and LTD Unit-* After updating a neuron, the spike from the neuron will be sent to LTP Unit. This signal decides whether the synaptic weight is "1" (on). LTD Unit begins to depress the synaptic weights depending on the LTD probability. A random 10-bit number $x$ will be generated by a 16-bit LFSR to compare with the LTD probability. If $x \leq$ *LTD probability*, the synaptic weight will be set to "0".

## 2.3 Overall Architecture of Wenquxing 22A

As mentioned above, we adopt our baseline, the NutShell processor, to support SNN computing. **Figure 2** shows the micro-architecture of Wenquxing 22A. The SNN Unit (shown as SNNU) is added to the pipeline, which is at the execution stage. Three stages in SNN workflow are handled by SPU (Spiking Process Unit), NU (Neuron Unit), and SU (Synapse Unit). All these components are integrated into SNNU. The SNN Special Register File is defined along with the General Registers File and ISU (Issue Unit) controls the instruction issuing, avoiding data hazards.

## 3 EXPERIMENT RESULTS

The MNIST data set [4][5] has 70,000 samples of handwriting digits from 0 to 9, among which 60,000 are used for training and 10,000 for testing. Every digital sample is a 28×28 gray-scale image with a maximum value of 255. We use Wenquxing 22A to classify this data set with the Binary Stochastic STDP learning rule.

### 3.1 Network Architecture

The architecture is presented in **Figure 3**. We design a simple 2-level supervised STDP-based SNN with pre-processing steps.

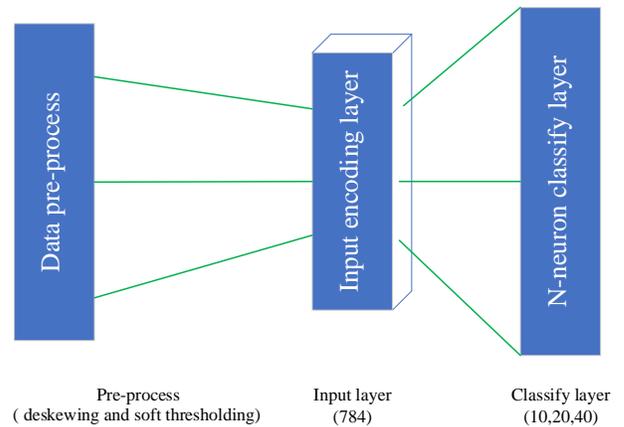

Pre-process (deskewing and soft thresholding) — Input layer (784) — Classify layer (10,20,40)

**Figure 3: Network Structure Applied to This Work.**

We set the number of synapses per neuron to be 28×28 to match the image format. The pre-processing steps applied to the MNIST data set are deskewing and soft thresholding, which are common practices for small networks to have better performance. To provide these samples to an SNN in Wenquxing 22A, a Poisson encoder is used to generate rate-based Poisson-distributed spikes to stimulate the input layer. this encoder converts the input data into a spike with the same shape. To generate the spike, we set a firing probability of a time cycle: $P = x$, where x needs to be normalized to [0,1]. All these pre-processing steps are executed in Wenquxing 22A.

**Table 1. MNIST Recognition Comparison Between Wenquxing 22A (this work) and Other SNN Systems Using STDP.**

| Name | Structure | Learning Rule | Resolution | Classification Accuracy | Encoding | Power Consumption | Hardware Utilization |
|---|---|---|---|---|---|---|---|
| Neftci 2014[12] | 784-500-40 | STDP + SNN-CD | 8-bit | 91.60% | Rate Poisson | N/A | N/A |
| | | | 5-bit | 89.40% | | | |
| **ODIN [14][14]** | **784-10** | **SDSP** | **3-bit** | **85.00%** | **Rate Poisson** | **25.949 W** | **63,411 LUT** |
| Yousefzadeh 2018 [13] | 784-6400-10 | STDP | 1-bit(64-bit classifier) | 95.70% | Rate Poisson | N/A | N/A |
| **This work** | **784-40** | **STDP** | **1-bit** | **91.91%** | **Rate Poisson** | **5.055 W** | **56,487 LUT** |

For the output layer shown in **Figure 3**, we choose single-layer fully connected networks of {10, 20, 40} LIF neurons. We set the on-chip training with an embedded binary stochastic STDP rule.

In the 10-neuron network, a teacher signal for supervised learning is used to correspond to the class of currently applied digit while other neurons are driven to a low firing activity. As for the networks consisting of more than 10 neurons, we applied the "Active learning" method. First, training images will be presented to 10 trained neurons as test samples. Therefore, we can get the error cases. Then those error samples will be presented as the training digits to new neurons supervised by the label of these images. Each neuron corresponds to a class of digits. Every training step will realize both test and training. Thus, based on test results of precious 10-neuron SNN, the rest extra neurons are trained with those digits that are not recognized successfully.

### 3.2 Comparison Results

According to recent works, we collect significant data of MNIST recognition from different spiking computing systems. **Table 1** shows the comparison of reported results on spiking versions of the static MNIST dataset [12][13][14]. "Structure" indicates the network structure. For instance, "784-10" means that the network has 2 layers and the number of neurons per layer is 784 and 10. "Learning Rule" is the training method for each work. "Resolution" is the precision of synaptic weights in each spiking system. "Classify Accuracy (CA)" indicates the test results of each work. We can see that in all systems utilizing lower than 8-bit STDP, Wenquxing 22A in this paper (represented by "this work", Bold in **Table 1**) does not differ much from their classification accuracy, or even better. To compare the power expenses, we realize our design and ODIN processor [14] (Bold in **Table 1**). We synthesize both Wenquxing 22A and ODIN to the same FPGA platform to make a comparison of power consumption between these two chips. The result is indicated in **Figure 4**.

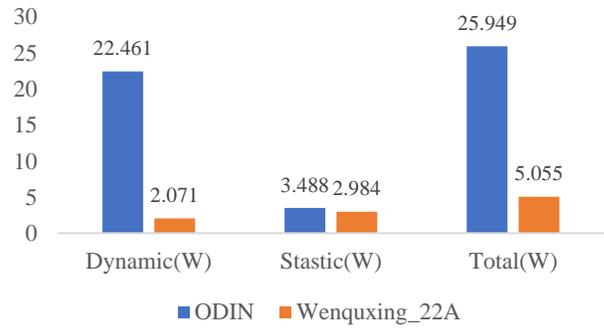

**Figure 4: Comparison of Power Consumption Between Wenquxing 22A and ODIN.**

We apply the same network architecture based on the recognition of the MNIST in [14], in which the images are compressed from 28×28 pixels to 16×16 pixels to fit the structure of synaptic weights from ODIN. From **Figure 4**, we can see the total power consumption is reduced from **25.949 W (ODIN)** to **5.055 W (Wenquxing 22A)**, which is detailed utilization of hardware resources illustrated in **Table 3** approximately 5 times less than the ODIN processor.

**Table 2. Hardware Utilization of Wenquxing 22A and ODIN.**

| Processor | Structure | LUT | FF | BRAM |
|---|---|---|---|---|
| ODIN | Nutshell + ODIN | 63411 (3.67%) | 75362 (2.18%) | 82.5 (3.07%) |
| This work | Wenquxing 22A | 56487 (3.27%) | 69702 (2.02%) | 73 (2.72%) |

According to[19], the ODIN processor is generated inside the Rocket Chip generated from Chipyard Repository [18], which is a control core for the ODIN processor. Since the Nutshell processor is the baseline of our design, we choose the Nutshell processor used as the control core for ODIN instead of the Rocket Chip in this comparison to make sure the difference between Wenquxing 22A and ODIN is only relevant to SNN computing. In this case, the hardware utilization report in **Table 2** indicates that Wenquxing consumes fewer hardware resources.

### 3.3 Wenquxing 22A Performance

We trained SNNs with varying numbers of neurons that are (10, 20, 40) neurons in binary stochastic STDP. The results of these networks were compared. **Figure 5** illustrated the Classification Accuracy (CA) of 10,000 MNIST test samples, which shows the maximum accuracy of binary weight SNNs ranging from 10, 20, and 40 neurons on Wenquxing 22A is 80.94%, 86.91%, and 91.91%. The increasing CA indicates that it is difficult to classify some digits with high similarity with small number of neurons. While the number of neurons in the output layer increases with the same $w_{exp}$, configurations more neurons can yield higher recognition accuracy.

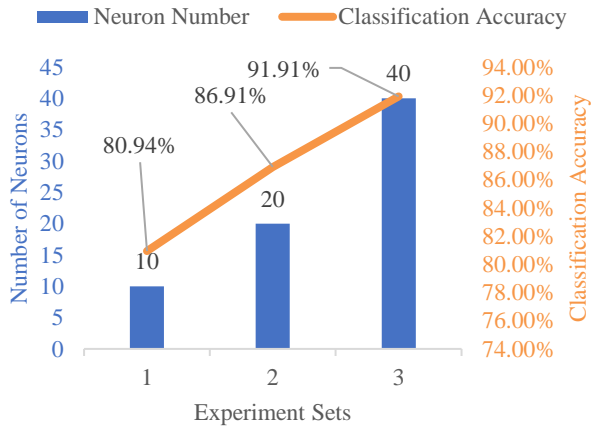

**Figure 5: Classification Accuracy of 3 Different Experiment Set (maximum CA 91.91%).**

However, we notice that as the number of neurons increases, there are chances that some neurons may turn to "death" (in 40-neuron SNN). Like the Dead ReLU Problem in ANN [11] we make a hypothesis that the appearance of this phenomenon may be caused due to the high learning rate. In this case, we later implemented the SNN with the same number of neurons but varying $w_{exp}$ parameters (128, 256, 512). This meta parameter can affect the number of effective synapses that ultimately remain by changing the "LTD probability". Since the learning rate in our network is controlled by LTD probability, adjusting of $w_{exp}$ parameter changes LTD probability and improves the situation of the "dead" neurons' appearance with higher CA in MNIST data set recognition.

## ACKNOWLEDGMENT

This work is supported by the Strategic Priority Research Program (Class A) of the Chinese Academy of Sciences titled Fundamental Software Ecosystem of RISC-V, by the Virtual Teaching and Research Section of the Ministry of Education of P.R. China for the Computer System and Processor Courses. We greatly appreciate the support and discussion from Dr. Yungang Bao, Mr. Huaqiang Wang and Mr. Zhijie Jia. We also appreciate Mr. Yang Liu, the senior Xilinx engineer, for supporting the FPGA platform.